\documentclass[journal]{IEEEtran}

\usepackage{cite,graphicx,amsmath,amssymb}
\usepackage{amsfonts,amssymb}
\usepackage{subfigure}
\usepackage{fancyhdr}
\usepackage{mdwmath}
\usepackage{mdwtab}
\usepackage{balance}
\usepackage{xcolor}
\usepackage{bm}
\usepackage{amsthm}
\usepackage{algorithm}
\usepackage{algorithmic}
\usepackage{multirow}
\usepackage{flafter}
\usepackage{mathrsfs}
\usepackage{url}
\usepackage{enumerate}
\usepackage{stfloats}

\usepackage[
top    = 1.70cm,
bottom = 0.82 in,
left   = 0.63 in,
right  = 0.63 in]{geometry}

\theoremstyle{definition}
\newtheorem{definition}{Definition}

\newtheorem{corollary}{Corollary}

\begin{document}

\title{Multi-agent Reinforcement Learning for Resource Allocation in IoT networks with Edge Computing}
\author{Xiaolan~Liu, Jiadong~Yu,
~Yue~Gao
\thanks{X. Liu and J. Yu  are with the Department of Electronic Engineering and Computer Science, Queen Mary University of London, London E1 4NS, U.K. (email: \{xiaolan.liu, jiadong.yu\}@qmul.ac.uk).}
\thanks{Y. Gao is with the Institute for Communication Systems, University of Surrey, GU2 7XH, U.K. (email: yue.gao@ieee.org)
}

 }

\maketitle

\begin{abstract}
To support popular Internet of Things (IoT) applications such as virtual reality, mobile games and wearable devices, edge computing provides a front-end distributed computing archetype of centralized cloud computing with low latency. However, it's challenging for end users to offload computation due to their massive requirements on spectrum and computation resources and frequent requests on Radio Access Technology (RAT). In this paper, we investigate computation offloading mechanism with resource allocation in IoT edge computing networks by formulating it as a stochastic game. Here, each end user is a learning agent observing its local environment to learn optimal decisions on either local computing or edge computing with the goal of minimizing long term system cost by choosing its transmit power level, RAT and sub-channel without knowing any information of the other end users. Therefore, a multi-agent reinforcement learning framework is developed to solve the stochastic game with a proposed independent learners based multi-agent Q-learning (IL-based MA-Q) algorithm.
Simulations demonstrate that the proposed IL-based MA-Q algorithm is feasible to solve the formulated problem and is more energy efficient without extra cost on channel estimation at the centralized gateway compared to the other two benchmark algorithms.
\end{abstract}


\section{Introduction}
The internet of things (IoT) networks are developed to provide different applications, such as mobile games, wearable and surveillance, etc., with the connected devices, like sensors and actuators, which leads to high demands for data processing\cite{bandyopadhyay2011internet}. The conventional idea is to transport the data bulks generated by a large number of end users to the cloud data centers for further analysis. However, moving such a large amount of data requires wide spectrum resource and the transmission delay can be prohibitively high\cite{pan2017future}. Although the devices with the ability to run artificial intelligence (AI) applications provide alternatives for processing data locally, they may suffer from poor performance and energy efficiency. Recently, fog computing has been proposed to provide a smooth data processing link from the end users to the cloud server with an extra layer, edge device, which removes the dilemma of data processing by moving it closer to the end users\cite{bonomi2012fog}.

In particular, edge computing can support latency-critical IoT applications and data processing in remote rural places. Edge devices are not like the cloud server that is believed to have infinite computation capacity and energy, they only can support limited number of end users for data processing\cite{resource_survey2018resource}.
Hence, cooperation between the edge device and the end users is necessary, and which can bring out a new era of data processing. Nevertheless, resource allocation, like computation, radio access technologies and spectrum resource, is essential to enhance energy and time efficiency in IoT edge computing networks.

\subsection{Related Work}
Fog computing, such as edge computing and cloudlet has been proposed as promising solutions to handle the large volume of security-critical and time-sensitive data\cite{fog_survey2018all,fog_survey12018internet}. Compared to using distant and centralized cloud data center resources, edge computing employed decentralized resources which were typically resource-constrained, heterogeneous and dynamic, thereby making resource management a critical challenge that needs to be addressed\cite{resource_survey2018resource,resource1_2018market}. To allocate the resources (e.g., computation, spectrum and radio access technologies), a novel market-based resource allocation framework with achieving high resource utilization in the system has been proposed in \cite{resource1_2018market}.

Recently, some methods have been explored to perform efficient resource allocation in edge computing networks, such as the computation offloading schemes have been proposed in mobile edge computing (MEC) networks \cite{MEC_mao2017joint,MEC_sardellitti2015joint,MEC_al2017energy}. A joint optimization of computation offloading scheduling and transmit power allocation scheme has been proposed in MEC system with single mobile user\cite{MEC_mao2017joint}, while joint optimizing radio and computational resources was considered when designing the optimal computation offloading scheme in multiuser MEC system\cite{MEC_sardellitti2015joint}.
\cite{MEC_al2017energy} adopted a time-division multiple access (TDMA) scheme and an orthogonal frequency-division multiple access (OFDMA) scheme to perform resource allocation. Moreover, exploring extra technologies used into MEC networks contributes to  more efficient computation offloading scheme, for instance, MEC networks with energy harvesting devices have larger ability to execute computation tasks, i.e., offload to edge server for execution or local execution \cite{EH_mao2016dynamic,EH1_wu2018wireless}, and millimeter wave communication technology used in MEC could significantly enhance the performance of the MEC infrastructure since it could increase the communication capacity of mobile back haul networks\cite{mmW_noghani2018generic}.

Moreover, the resource allocation problem becomes aggravated when considering the randomness and dynamics of wireless networks, such as users' mobility, uncertain channel quality, random task-arrival and energy resources (if considering energy harvesting). Hence, the conventional optimization methods, like Lyapunov optimization and convex optimization techniques\cite{Lyapunov_he2018energy}, cannot address these challenges to obtain optimal computation computation offloading schemes. Machine learning is a promising and powerful tool to provide autonomous and effective solutions in an intelligent manner to enhance the edge computing networks \cite{edge_zhu2018towards,edge1_du2018fast}. Note that exploring efficient resource allocation can be achieved by designing computation offloading scheme, which basically can be modeled as a Markov decision process (MDP). It could be solved by a classical single-agent Q-learning algorithm, to break the curse of dimensionality, deep learning is used to approximate the Q-value function in Q-learning. \cite{RL1_chen2018performance,RL_zhang2018deep} adopted deep Q-network (DQN) to learn the optimal policy by solving the formulated computation offloading problem in MEC systems. Moreover, in our previous work, we investigated resource allocation in IoT networks with edge computing, and proposed computation offloading algorithm based on Q-learning and DQN to obtain the optimal computation offloading policy by considering single agent scenario\cite{liu2019resource,liu2020resource}.

\subsection{Motivation and Contributions}
As mentioned above, machine learning has played an important role in learning optimal policies for computation offloading problem. However, most research works were focused on dynamic computation offloading framework for single-user case\cite{RL1_chen2018performance,RL_zhang2018deep,e_sun2018deep}.
They take a representative user in the MEC system as an example to design the optimal computation offloading scheme without considering the conferences from other users. In this case, the complete network information is assumed to be known at the user, and then transmit power and radio access technology allocation are performed. Besides, other existing works focus on centralized approaches, where a centralized network controller was used to support data training and resource management\cite{e_yu2017computation}. Nevertheless, with the network scale increasing, it's more challenging for the network controller to deal with modelling and a large amount of computational tasks. Multi-agent reinforcement learning (MARL) is able to provide efficient resource management for edge computing IoT networks in a distributed aspect especially when end users only can observe their local information.

MARL is focused on models including multiple agents that learn dynamically interacting with their environment. Compared to the single-agent scenario that the environment changes its state only based on the action of an agent, the new environment state depends on the joint action of all the agents in MARL scenario. The MARL has a few benefits: 1) Agents learn their strategies in distributed manners with observing local environment information; 2) Agents can share experience by communicating with each other; 3) MARL is more robust if some agents fail in the multi-agent system, the rest agents can take over some of their tasks. The MARL was popularly adopted to perform resource allocation in different wireless communication networks\cite{m1_chen2012stochastic,m2_ranadheera2017mobile,m3_sharma2019multi,m4_nasir2019multi,m5_cui2019multi}.
In \cite{m1_chen2012stochastic}, a multi-agent reinforcement learning approach was proposed to explore stochastic power adaption in cognitive wireless networks. Multi-agent deep reinforcement learning was adopted to solve dynamic power allocation problem in wireless networks\cite{m3_sharma2019multi,m4_nasir2019multi}, here, deep reinforcement learning was used to learn the optimal power transmission strategies distributively. A multi-agent dynamic resource allocation algorithm has been proposed for multi-UAV downlink networks to jointly design user, power level and subchannel selection strategies\cite{m5_cui2019multi}. Moreover, in \cite{m2_ranadheera2017mobile}, game theory and reinforcement learning were applied to address efficient distributed resource management in MEC systems.

Therefore, invoking MARL to edge computing networks provides promising solutions for intelligent resource management with distributed learning. In IoT edge computing networks, due to its characteristics of ultra-high density, distributed nature, random channel conditions as well as resource-hungry and delay-constrained, to the best of our knowledge, multi-user computation offloading problems haven't been well investigated. In this paper, motivated by the applications of MARL in resource management, we use a stochastic game to formulate multi-user computation offloading problem in IoT edge computing networks. Moreover, an independent learners based multi-agent Q-learning (IL-based MA-Q) algorithm is proposed to solve the formulated problem. Specifically, we assume that all the end users decide if to offload their computation tasks to the edge server distributively without the assistance of a central controller and each user can observe its local environment information.
The major contributions of this paper are presented as:
\begin{enumerate}
\item We investigate computation offloading with resource allocation problem in multi-user IoT edge computing networks by jointly selecting the transmit power level, the sub-channel and the radio access technology.
\item We use a stochastic game to formulate the multi-user computation offloading with resource allocation problem in IoT edge computing networks, in which each user becomes a learning agent to learn the optimal computation offloading policy with each resource allocation solution as an action taken by end users.
\item We propose an IL-based MA-Q computation offloading algorithm to explore the optimal computation offloading policy with efficient resource allocation solutions for end users. Here, each agent runs an independent learning algorithm with considering other users as part of the environment, and there is no communication among users.
\item Numerical results demonstrate that the proposed IL-based MA-Q computation offloading algorithm is superior to the centralized and random computation offloading schemes, moreover, it's more energy efficient without extra cost on channel estimation.
\end{enumerate}

\subsection{Organization}
The remainder of this paper is organized as follows. In Section II, we present a system model for a multi-user IoT edge computing network. The computation offloading with resource allocation problem in the considered IoT network is formulated with an Markov game in Section III. Section IV proposes an IL-
based MA-Q computation offloading algorithm to solve the formulated problem. Simulations are presented in Section V, finally the conclusions are drawn in Section VI.

\section{System Model}
\begin{figure}[t]
  \centering
  \includegraphics[width=3.5in]{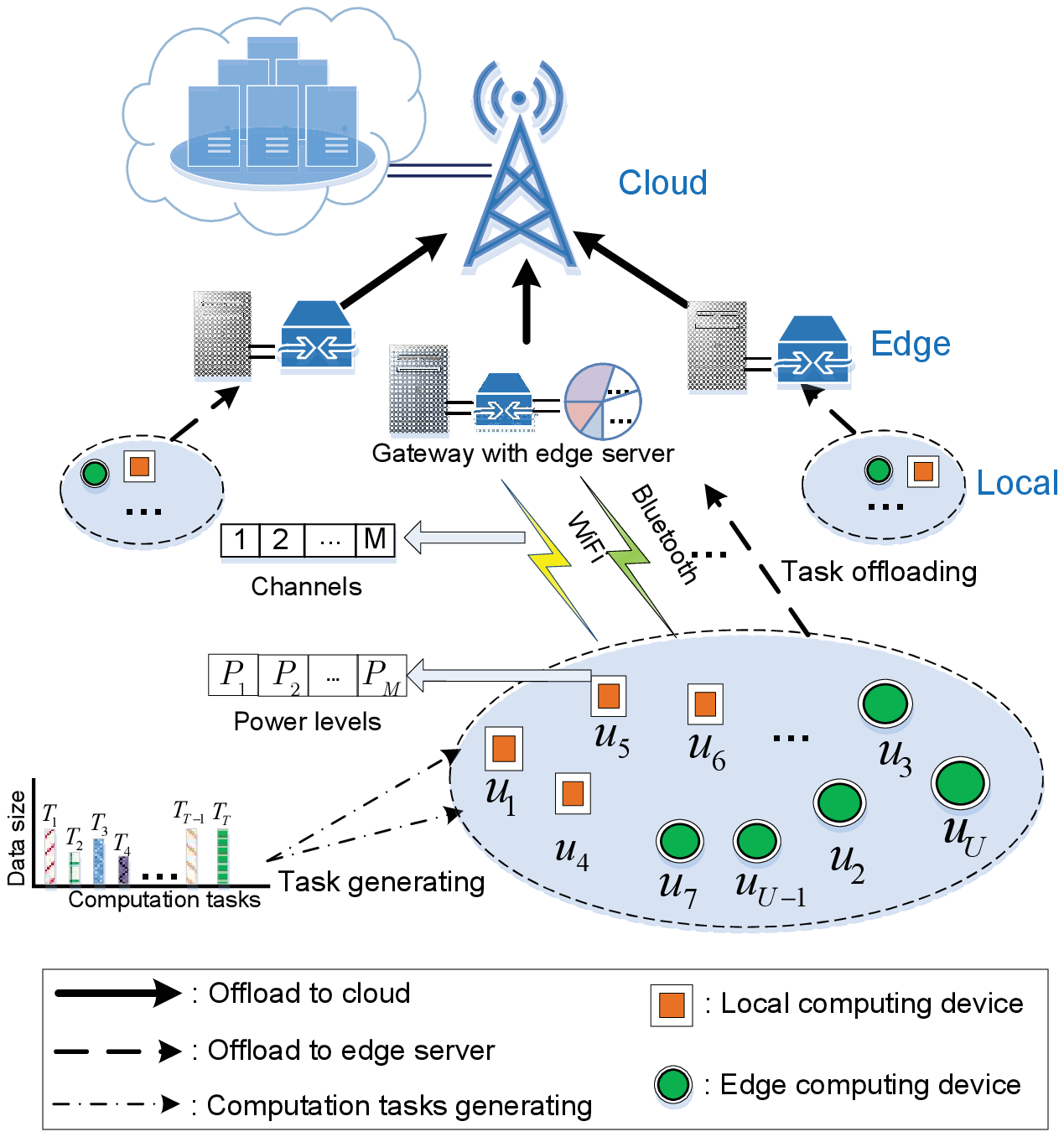}\\
  \caption{Computation offloading model in IoT networks.}
  \label{system model}
\end{figure}

As shown in Fig.~\ref{system model}, an IoT framework is consisted of various IoT networks with three layers including cloud platform, edge and local end devices. We consider a multi-user IoT edge computing network, which includes $U$ single antenna end users and one single antenna edge server. Here, the end users could be different kinds of sensors or mobile devices which can process some small computation tasks, while the edge device could be the gateway and the access point that has much higher computation capacity to provide computation tasks processing for the end users.
We assume the end users are stationary and randomly distributed around the gateway, they sense data and process it with two possible ways: 1) local computing: process computation tasks locally at end user; 2) edge computing: offload computation tasks to the gateway and process them at the edge server. The end users have limited computation capacity and the generated computation tasks are latency constrained, so they decide to offload some computation tasks to the gateway by choosing the possible radio access technologies denoted by $\mathcal {{RA}}=\{{RA}_{1},...,{RA}_{N}\}$ with each radio access technology has $M$ orthogonal sub channels, denoted by $\mathcal{CH}=\{{CH}_1,...,{CH}_M\}$. Note that each radio access technology is operated on different radio frequencies. From Fig.~\ref{system model}, different computation tasks are continuously generated at each end user. The gateway in the considered IoT edge computing network serves a set $\mathcal{U}=\{u_1,...,u_U\}$ of end users. Each end user $u_i$ generates computation tasks with the $j^{th}$ task denoted as $T_{i,j}(d_{i,j}, L_{i,j})$, where $d_{i,j}$ is the task size and $L_{i,j}$ is the delay constraint on task execution.

This paper is focused on dynamic computation offloading with efficient resource allocation in the considered IoT edge computing network by selecting transmit power levels, radio access technologies and sub-channels without assistance of the centralized controller. Here, each end user only can observe its local information, such as the channel state information (CSI) between the end user and the gateway and feedback from the gateway. The system is assumed to operate on a time slotted structure and we discrete the time horizon into slots, indexed by an integer $k \in \mathcal {K} = \{ 1,2,...,K\} $. At each time slot $k$, each end user makes its own decisions on computation tasks offloading distributively with their local observed information.

\subsection{Task Execution Model}
In the considered multi-user IoT edge computing network, each end user is possible to execute its tasks locally or offload to the edge server and perform task execution there. In this section, the detailed task execution model in both cases is built.

\subsubsection{Local Computing Model}
Let the end user $u_i$ choose to execute its task locally in time slot $k$. $\nu$ presents the number of CPU cycles required to process 1 bit data, and $e_i^L$ denotes the computing energy consumption of each CPU cycle at the end user. Hence, the computing energy consumption of task $T_{i,j}$ is calculated as
\begin{equation}\label{local_power}
E_{i,j}^L=d_{i,j} \cdot \nu \cdot e_i^L,
\end{equation}
Moreover, let $f_i$ indicate the computation capacity of the end user $u_i$. Then the time delay of local task execution is
\begin{equation}\label{local_delay}
D_{i,j}^L=d_{i,j} \cdot \nu/f_i.
\end{equation}

\subsubsection{Edge Computing Model}
Compared to the local device, edge server has much more powerful computation capacity $f$ and more stable power supply. Then the computation time of the task execution at edge server is
\begin{equation}\label{edge_delay}
D_{i,j}^E=d_{i,j} \cdot \nu/f.
\end{equation}
Similarly, let $e^E$ denote the computing energy consumption of the edge server. Hence, the energy consumption by processing the offloaded computation task $T_{i,j}$ is calculated as
\begin{equation}\label{edge_power}
E_{i,j}^E=d_{i,j} \cdot \nu \cdot e^E,
\end{equation}

\subsection{Task offloading Model}
As mentioned before, end users can offload their computation tasks to the gateway, and the edge server is able to support efficient computation task execution with its powerful computation ability. Assuming that the channels between each end user and the gateway follows Rayleigh fading distribution, each user can select one radio access technology from $\mathcal {{RA}}$ and one sub-channel from $\mathcal{CH}$ to transmit data to the gateway. The transmission rate can be presented as
\begin{equation}\label{transmission rate}
R_{i,j}^{m,n} = {B^{m,n}}{\log _2}(1 + \frac{{P_i^T{h_i^{m,n}}}}{{{\sigma ^2} + \sum\limits_{z = 1,{\kern 1pt} z \ne i{\kern 1pt} }^Z {\delta _z^{m,n}{h^{z \to i}}P_z^T} }})
\end{equation}
where $P_i^T$ is the transmit power of the user $u_i$, $h_{m,n}$ is the channel gain of radio access technology ${RA}_n$ in sub-channel ${CH}_m$, and $h^{z \to i}, z \in \mathcal {Z}$ is the channel gain from other users to user $u_i$. The noise is assumed to be white Gaussian channel noise with its variance as $\sigma ^2$. $P_z^T$ is the interferences from the other end users $\mathcal {Z}=\{1,...,Z\}$ choose the same channel ${CH}_{m,n}$ \footnote{For simplicity, ${CH}_{m,n}$ is used to indicate the channel ${CH}_m$ with radio access technology ${RA}_n$ in the rest text.}, and $\delta _z^{m,n}$ indicates if the user $u_z$ take up the same channel or not, which is defined as
\begin{equation}\label{alpha}
\delta _z^{m,n} = \left\{ {\begin{array}{*{20}{c}}
{1,{\kern 1pt} {\kern 1pt} {\kern 1pt} {\kern 1pt} {\kern 1pt} {u_z}{\kern 1pt} {\kern 1pt} {\kern 1pt} {\kern 1pt} in{\kern 1pt} {\kern 1pt} {\kern 1pt} {CH}_{m,n}{\kern 1pt} }\\
{0,{\kern 1pt} {\kern 1pt} {\kern 1pt} {\kern 1pt} {\kern 1pt} {\kern 1pt} {\kern 1pt} otherwise{\kern 1pt} {\kern 1pt} {\kern 1pt} {\kern 1pt} {\kern 1pt} {\kern 1pt} {\kern 1pt} }
\end{array}} \right.
\end{equation}

Then the transmission delay of the task $T_{i,j}$ offloaded from user $u_i$ to the gateway is given as
\begin{equation}\label{transmission_delay}
D_{i,j}^T = {d_{i,j}}/R_{i,j}^{m,n}
\end{equation}
Moreover, the consumed energy for computation task offloading is calculated as
\begin{equation}\label{transmission_power}
E_{i,j}^T = P_i^T \cdot D_{i,j}^T.
\end{equation}

Here, the observed signal-to-interference-plus-noise ratio (SINR) from user $u_i$ to the gateway over channel ${CH}_{m,n}$ with offloading task $T_{i,j}$ is given as
\begin{equation}\label{SINR}
\gamma_{i,j}^{m,n}=\frac{{P_i^T{h^{m,n}}}}{{{\sigma ^2} + \sum\limits_{z = 1,{\kern 1pt} z \ne i{\kern 1pt} }^Z {\delta _z^{m,n}{h^{z \to i}}P_z^T} }}
\end{equation}

In this paper, each end user adopts discrete transmit power control, with the transmit power values expressed as a vector $\mathcal {P}_i^T=\{P_1,...,P_X\}$. At time slot $k$, each end user selects its transmit power $P_{i,x}^T, \forall x=\{1,...,X\}$ from $\mathcal {P}_i^T$ when it chooses to offload its computation task, otherwise, $P_{i,x}^T=0$ indicates the user chooses to execute its task locally. Hence, we define a finite set to present the possible transmit power levels selected by the end user $u_i$ as
\begin{equation}\label{transmit power set}
\mathcal {P}_i^T=\{P_0,P_1,...,P_X\}, P_x \ne 0,\forall i \in \mathcal {U}
\end{equation}
where $P_0=0$ indicates the user chooses local execution.
Similarly, the possible radio access technologies and possible sub-channels under each radio access technology can be chosen by the end user $u_i$, which are defined as finite sets, respectively.
\begin{equation}\label{radio access}
\begin{array}{l}
\mathcal {RA}_i=\{{RA}_1,{RA}_2,...,{RA}_N\}, \forall i \in \mathcal {U} \\
\mathcal {CH}_i=\{{CH}_1,{CH}_2,...,{CH}_M\}, \forall i \in \mathcal {U}
\end{array}
\end{equation}

\begin{figure}[t]
  \centering
  \includegraphics[width=3.5in]{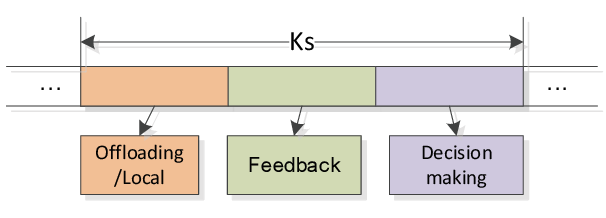}\\
  \caption{The timeslot structure of the computation offloading problem.}
  \label{timeslot_structure}
\end{figure}

The considered multi-user IoT edge computing network is assumed to operate on discrete time horizon with each time slot equal and non-overlapping. Moreover, we assume the communication parameters keep unchanged during each time slot. Each time slot is denoted by $k$, and it can last $K_s$ duration, the time slot structure is shown in Fig.\ref{timeslot_structure}. During the time slot $k$, each end user executes its computation tasks according to the computation decisions made in the last time slot $k-1$, and then receives some feedbacks from the gateway, finally it has to make decision on task execution by the end of time slot $k$, i.e., $k+1=k+K_s$.

\section{Problem Formulation with Markov Game}
In this section, we formulate the computation offloading with resource allocation problem first, and then we use Markov game to model the formulated problem of joint choosing task execution decision, transmit power level, radio access technology and sub-channel by multiple users in IoT edge computing networks.

\subsection{Computation Offloading Problem Formulation}
In IoT edge computing networks, energy consumption and time delay are two main concerns while designing efficient computation offloading policy.
If the end user chooses to offload its computation task, it will take up spectrum and computation resource of the network, which, in turn, reduces the resource that other users can be allocated. Moreover, from (\ref{SINR}), larger transmit power means higher transmission rate, smaller transmission delay, but causes more interference to other end users. Therefore, it's necessary to design optimal joint computation task offloading scheme and with efficient resource allocation among end users. Here, we consider the system cost, defined as the weighted sum of energy consumption and task execution delay, as an index to evaluate the decisions on task offloading and resource allocation made by end users, given by
\begin{equation}\label{system_cost}
C_i = {{E_i} + \beta {D_i}}
\end{equation}
where $E_i$ is the energy consumption of user $u_i$, and $D_i$ is the task execution delay. Specifically, the system cost is considered as a negative reward function in our problem, which can indicate what are good and bad decisions for the agents.

At any time slot $k$, the end user $u_i$ can be at two possible states: executing task locally or transmitting task to the gateway. Based on the states, each end user selects transmit power level, radio access technology and sub-channel. Then it receives a reward to evaluate the performance of the selected actions. Hence, the design of the reward function directly guides the learning process. In this paper, we design a segmented function to present the reward function $r_i(k)$ of end user $u_i$
\begin{equation}\label{reward}
{r_i}(k) = \left\{ {\begin{array}{*{20}{c}}
\begin{array}{l}
{C_{i,j}^L,\;\;\;\; \gamma _{i,j}^{m,n} = 0}\\
C_{i,j}^E +C_{i,j}^T, \;\;\;\;\;\;\;\;\gamma _{i,j}^{m,n} > \bar \gamma^n ,\; {W_{all}} < \bar W\\
C_{i,j}^E +C_{i,j}^T+\omega ,\;\; \gamma _{i,j}^{m,n} > \bar \gamma^n ,\; {\kern 1pt} {W_{all}} > \bar W{\kern 1pt}\\
{C_{i,j}^T + \varpi ,\;\; \gamma _{i,j}^{m,n} <  \bar \gamma }
\end{array}\\
\end{array}} \right.
\end{equation}
where $\bar \gamma^n$ denotes the threshold of the SINR at the gateway, which is assumed to be the same for the end users with the same radio access technology. $W_{all}$ indicates the required computation capacity of all the offloaded tasks, while $\bar W$ is the computation capacity of the edge server. $\omega$ indicates the waiting cost caused by non-enough computation capacity of the edge server. $\varpi$ is the penalty for failed computation task transmission. $C_{i,j}^L$, $C_{i,j}^E$, and $C_{i,j}^T$ presents the cost of local task execution, edge task execution and task transmission, they are calculated as
\begin{equation}\label{cost_LET}
\begin{array}{l}
C_{i,j}^L={E_{i,j}^L + \beta D_{i,j}^L}\\
C_{i,j}^E={E_{i,j}^E + \beta D_{i,j}^E}\\
C_{i,j}^T={E_{i,j}^T + \beta D_{i,j}^T}
\end{array}
\end{equation}

Note that the instant reward in any time slot $k$ of the end user $u_i$ mainly relies on its observed information. 1) the observed information: the taken actions including the transmit power level $P_i^T(k)$, the radio access technology $RA_i(k)$ and the subchannel $CH_i(k)$, and it relates to the current channel gain $h_i^{m,n}(k)$ and the remaining computation capacity of the edge server. 2) the unobserved information: the actions taken by other end users in the same IoT network and the channel gains.

Next, we consider to maximize the long-term reward $v_i(k)$ by selecting the transmit power, radio access technology and sub-channel at each time slot, which is given by
\begin{equation}\label{long_term_reward}
v_i(k)=\sum\limits_{\tau  = 0}^{ + \infty } {{\lambda ^\tau }{r_i}(k + \tau  + 1)} ,
\end{equation}
where $\lambda \in [0,1]$ is the discount factor. $v_i(k)$ presents the discounted sum of future rewards, which can be used to measure the taken action by end user $u_i$.

The action space of each end user contains the set of possible transmit level $\mathcal {P}_i^T$, radio access technologies $\mathcal {RA}_i$ and sub-channels $\mathcal {CH}_i$, which can be denoted as $\mathcal {A}_i= \mathcal {P}_i^T \times \mathcal {RA}_i \times \mathcal {CH}_i$. Moreover, at any time slot $k$, the goal of each end user is to take an optimal action $a_i^*(k)=({P}_i^{T*},{RA}_i^*,{CH}_i^*) \in \mathcal {A}_i$ with maximizing the long term reward in  (\ref{long_term_reward}). Specifically, we consider the cost as the reward function, which is a negative reward, so we have to minimize the long-term reward here.  Therefore, the optimization computation offloading problem of end user $u_i$ can be formulated as
\begin{equation}\label{optimization_problem}
a_i^*(k) = \mathop {\arg \min }\limits_{{a_i} \in {\mathcal {A}_i}} {v_i}(k)
\end{equation}

However, the design of computation offloading scheme for the considered multi-user IoT edge computing network consists of $U$ subproblems as mentioned above, which corresponds to $U$ end users. Moreover, each end user has no information of other end users and it's impossible and hard to share their own information, like the taken actions, received rewards, to each other. We cannot find a direct tool to solve this problem accurately, so we first model this optimization problem with a non-cooperative stochastic game, and then propose a multi-agent reinforcement learning framework to solve it.

\subsection{Stochastic Game}
In this subsection, we consider using stochastic game framework to model the design of multi-user computation offloading schemes while minimizing the system cost. In the considered multi-user IoT edge computing network, each end user is considered as an agent to learn its optimal computation offloading policy by interacting with the environment. This is an N-agent game, in which each end user learns its computation offloading policy without any cooperation with the other end users. Each agent observes the environment state $s_i(k)\in \mathcal{S}_i$, then independently takes one action $a_i(k)\in \mathcal{A}_i$, choosing its transmit power, radio access technology and sub-channel. Consequently, each agent receives a reward $r_i(k)=r_i(s_i(k), a_1(k),...a_U(k) )$, i.e., the system cost, and transits to a new state $s_i(k+1)\in \mathcal{S}_i$ depending on the actions of all the involved agents.

A stochastic game is the  generalization of Markov decision process in multi-agent case, also named as a Markov game, which is denoted by a tuple $<\mathcal {S}, U, \mathcal{A}, \mathbb {P}, \mathcal{R}>$.
\begin{itemize}
\item $\mathcal {S}$ is the environment states that include the state of each player, $\mathcal {S}=\mathcal {S}_1 \times \mathcal {S}_2 \cdots \times \mathcal {S}_U$;
\item $U$ is the player number.
\item $\mathcal{A}$ is the joint action set $\mathcal{A}=\mathcal{A}_1 \times \mathcal{A}_2 \cdots \times \mathcal{A}_U$,
\item $ \mathbb {P}: \mathcal{S} \times \mathcal{A} \times \mathcal{S} \in [0,1]$ is the state transition probability function
\item $\mathcal{R}=\{R_1,\cdots R_U\}$ contains all the reward functions of the agents.
\end{itemize}
\subsection{Computation Offloading Game Formulation}
Based on the definition of the stochastic game, we formulate our multi-user computation offloading problem as a stochastic game by figuring out each item in the tuple.

\subsubsection{\textbf{Action}}
In the considered multi-user IoT edge computing network, each end user $u_i$ is considered as an intelligent agent, at any time slot $k$, it takes an action including selecting transmit power level $P_i^T(k)$, radio access technology ${RA}_i(k)$ and sub-channel ${CH}_i(k)$ to complete task execution. We designate $a_i(k) \in \mathcal{A}_i=\mathcal {P}_i^T \times \mathcal {RA}_i \times \mathcal {CH}_i$ as the end user $u_i$'s action at time slot $k$. Hence, the action space of the task offloading game is $\mathcal {A}=\prod_{i \in \mathcal { {U}}} {\mathcal {A}_i} $.

\subsubsection{\textbf{State}}
There is no cooperation among the competing end users, so the environment state is defined based on each end user's local observations. At time slot $k$, the state observed by the end user $u_i$ is given by
\begin{equation}\label{state}
s_i(k)=(\mathcal {L}_i(k), \mathcal {I}_i(k),\mathcal {J}_i(k))
\end{equation}
where $\mathcal {L}_i(k) \in \{0,1\}$ indicates whether the transmit power $P_i^T(k)$ of the user $u_i$ is equal to zero or not, denoted as
\begin{equation}\label{power_state}
\mathcal {L}_i(k) = \left\{ {\begin{array}{*{20}{c}}
\begin{array}{l}
{0, if \; P_i^T(k)=P_0=0}\\
{1, \;otherwise}
\end{array}
\end{array}} \right.
\end{equation}
$\mathcal {I}_i(k) \in \{0,1\}$ indicates whether the $u_i$'s computation offloading can be recognized by the gateway, that is, the received SINR $\gamma_{i,j}^{m,n}(k)$ of user $u_i$ is above or below its threshold $\bar \gamma^n$, i.e.,
\begin{equation}\label{SINR_state}
\mathcal {I}_i(k) = \left\{ {\begin{array}{*{20}{c}}
\begin{array}{l}
{1, if \; \gamma_{i,j}^{m,n}(a_i(k), \mathbf {a}_{-i}(k))(k) > \bar \gamma^n }\\
{0, \;otherwise}
\end{array}
\end{array}} \right.
\end{equation}
where $\mathbf {a}_{-i}(k)=(a_1(k),...,a_{i-1}(k),a_{i+1}(k),...,a_{U}(k)) \in \mathcal {A}_{-i}=\mathcal {A}_{1} \times ... \times \mathcal {A}_{i-1} \times \mathcal {A}_{i+1} \times ... \times \mathcal {A}_{U}$ indicates the action vector of the other end users.

Moreover, $\mathcal {J}_i(k)\in \{0,1\}$ denotes the broadcast information from the gateway that presents if the gateway's computation capacity is enough to support all the computation tasks from offloading users, given by
\begin{equation}\label{computation_state}
\mathcal {J}_i(k) = \left\{ {\begin{array}{*{20}{c}}
\begin{array}{l}
{1, if \; {W_{all}} \leq \bar W}\\
{0, \;otherwise}
\end{array}
\end{array}} \right.
\end{equation}

\subsubsection{\textbf{Reward}}
The reward $r_i(s_i(k),a_i(k),\mathbf {a}_{-i}(k))$ of end user $u_i$ in state $s_i(k)$ presents the immediate return by $u_i$ taking action $a_i(k)$ while the other end users taking actions $\mathbf {a}_{-i}(k)$ at time slot $k$. It's rewritten as
\begin{equation}\label{reward_r}
\begin{array}{l}
r(s_i(k),a_i(k),\mathbf {a}_{-i}(k))\\
= \left\{ {\begin{array}{*{20}{c}}
\begin{array}{l}
{C_{i,j}^L,\;\;\;\; if \;\mathcal {L}_i(k) = 0}\\
C_{i,j}^E +C_{i,j}^T, \;\;\;\;\;\;if \; (\mathcal {L}_i(k), \mathcal {I}_i(k),\mathcal {J}_i(k))=(1,1,1)\\
C_{i,j}^E +C_{i,j}^T+\omega ,\;\;if \; (\mathcal {L}_i(k), \mathcal {I}_i(k),\mathcal {J}_i(k))=(1,1,0)\\
{C_{i,j}^T + \varpi ,\;\; if \; \mathcal {L}_i(k) = 1, \mathcal {I}_i(k)=0 }
\end{array}\\
\end{array}} \right.
\end{array}
\end{equation}

Recall that a policy, $\pi_i(s_i,a_i)$, is a mapping from each state, $s_i \in \mathcal {S}_i$ and $a_i \in \mathcal {A}_i$ to a probability $\pi_i(s_i,a_i)=\Pr(a_{t}=a\mid s_{t}=s) \in [0,1]$. It gives the probability of taking action $a_i$ in state $s_i$. Specifically, for end user $u_i$ in the state $s_i$, it has mixed strategy as $\pi_i(s_i)=\{\pi_i(s_i,a_i) \mid  a_i \in \mathcal {A}_i\}$. Hence, in a stochastic game, a joint strategy for $U$ players is defined a strategy vector $\pi =(\pi_1(s_1), \pi_2(s_2),...,\pi_U(s_U))$ with each strategy belonging to each player. Based on the probabilistic policies, we formulate the optimization goal of each end user $u_i$ in (\ref{long_term_reward}) into its discounted expected form as
\begin{equation}\label{expected_reward}
{V_i}({s_i},{\pi _i}) = E[\sum\limits_{\tau  = 0}^{ + \infty } {{\lambda ^\tau }{r_i}(k + \tau )} \mid s_i(k)=s_i,\pi_i]
\end{equation}
where $E[\cdot]$ is the expectation operation, which calculates the station transition under strategy $\pi_i$ in state $s_i$.
The state transition from environment state $s_i(k)$ to new state $s_i(k+1)$ is determined by the joint strategy of all the end users. Moreover, in the non-cooperative game, at each time slot $k$, each end user $u_i$ chooses its strategy $\pi_i(s_i)(k)$ independently in state $s_i(k)$ to maximize its discounted reward ${V_i}({s_i},{\pi _i})$, and then it receives its current individual reward based on the joint strategy $\pi$. Here, the goal of each end user is to learn the optimal strategy $\pi_i^*$ from any state $s_i \in \mathcal {S}_i$, i.e., the optimal strategies of other end users are learned as $\bm{\pi}_{-i}^*=(\pi_1^*,...,\pi_{i-1}^*,\pi_{i+1}^*,...,\pi_U^*)$.
Hence, the expected reward is reformulated as in (\ref{re_expected_reward}).
\begin{figure*}
\begin{equation}\label{re_expected_reward}
\begin{array}{l}
{V_i}({s_i},{\pi _i},{\bm {\pi}  _{-i}}) = E[\sum\limits_{\tau  = 0}^{ + \infty } {{\lambda ^\tau }{r_i}({s_i(k + \tau )},{\pi _i(s_i(k + \tau ))},{\bm {\pi} _{-i}(s_i(k + \tau  ))})} \mid s_i(k)=s_i]\\
\bm {\pi} _{-i}(s_i(k ))=(\pi_1(s_1(k)),..., \pi_{i-1}(s_{i-1}(k)),\pi_{i+1}(s_{i+1}(k)),....\pi_U(s_U(k)))
\end{array}
\end{equation}
\end{figure*}

\begin{definition}\label{Nash-E}
\emph{A Nash equilibrium is a set of $U$ optimal strategies $(\pi_1^*,...,\pi_U^*)$, in which no end user can receive any higher reward by changing only their own strategies. That is, for each end user $u_i \in \mathcal {U}$ in each state $s_i \in \mathcal {S}_i$,
\begin{equation}\label{Nash}
V_i(s_i,\pi_i^*,\bm {\pi}_{-i}^*) \geq V_i(s_i,\pi_i,\bm {\pi}_{-i}^*),\; \forall \pi_i \in \Pi_i,
\end{equation}
where $\Pi_i$ is the set of possible strategies can be taken by end user $u_i$.}
\end{definition}

\begin{definition}\label{NE-exist}
\emph{1) Every finite stochastic game has a Nash equilibrium if it has a finite number of players, $U$, finite action set, $\mathcal {A}$, and set of states, $\mathcal {S}$.  \\
2) A game with infinitely many stages if the total payoff is the discounted sum, ${V_i}({s_i},{\pi _i},{\bm {\pi}  _{-i}})$.}
\end{definition}
This means, there always exists a NE in our formulated computation offloading game. In a NE, each end user has gotten its optimal strategy, no end user can get any better strategies by changing only their own strategies. Therefore, in this paper, each end user $u_i$ is aim to find a NE strategy for any state $s_i$.
\begin{figure}[t]
  \centering
  \includegraphics[width=3.3in]{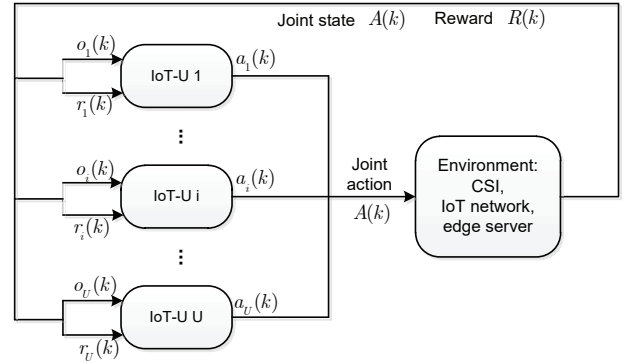}\\
  \caption{The MARL framwork for multi-user IoT edge computing network.}
  \label{MARL}
\end{figure}

\section{Proposed Multi-Agent Reinforcement Learning Algorithm}
In this section, we first describe computation offloading in multi-user IoT network with multi-agent reinforcement learning framework. Then multi-agent Q-learning based computation offloading algorithm is proposed to address the formulated computation offloading game.

\subsection{MARL Framework for Multi-user IoT edge computing network}
Fig. \ref{MARL} illustrates an MARL framework for an multi-user IoT edge computing network. Here, in time slot $k$, each end user $u_i$ at state $s_i(k)$ takes the action $a_i(k) \in \pi_i(s_i)$, and then the environment is changed to a new joint state $S(k+1)=\{s_1(k+1),...,s_i(k+1),...,s_U(k+1)\}$ based on the joint action $A(k)=\{a_1(k),...,a_i(k),...,a_U(k)\}$ taken by all the end users. Finally, each end user observes its local information $o_i(k)$ and gets its own reward $r_i(k)$. Since the future state $S(k+1)$ only depends upon the present state $S(k)$ and the taken action $A(k)$, this dynamic MARL process has Markov property such that it's formulated into a Markov game, i.e., stochastic game. Specifically, the stochastic game with a single player is modelled as a Markov decision process (MDP), moreover, the decision problem faced by a player in a stochastic game when all other players choose a fixed profile of stationary strategies is equivalent to an MDP \cite{neyman2003markov}.

In the non-cooperative game, each end user $u_i$ chooses the strategy $\pi_i(s_i)$ independently to maximize its total expected discounted reward, defined as the value function, from (\ref{re_expected_reward}). The value function can be decomposed into two parts as shown in the Bellman equation:
\begin{equation}\label{value-Bellman}
\begin{array}{l}
{V_i}({s_i},{\pi _i},{\bm {\pi}  _{-i}})\\
= E[{r_i}({s_i},{\pi _i(s_i)},{\bm {\pi} _{-i}(s_i)})\\
+ \sum\limits_{\tau  = 1}^{ + \infty } {{\lambda ^\tau }{r_i}({s_i^ \prime},{\pi _i(s_i^ \prime)},{\bm {\pi} _{-i}(s_i^ \prime)})} \mid s_i(k)=s_i],
\end{array}
\end{equation}
where we let $s_i(k)=s_i$ and $s_i(k+1)=s_i^\prime$, so the Bellman equation is formulated as
\begin{equation}\label{re-value-Bellman}
\begin{array}{l}
{V_i}({s_i},{\pi _i},{\bm {\pi}  _{-i}}) \\
= E[{r_i}({s_i},{\pi _i(s_i)},{\bm {\pi} _{-i}(s_i)})
+ \lambda {V_i}({s_i^\prime},{\pi _i},{\bm {\pi}  _{-i}}) \mid s_i(k)=s_i]\\
=E[{r_i}({s_i},{\pi _i(s_i)},{\bm {\pi} _{-i}(s_i)})]+\lambda E[ {V_i}({s_i^\prime},{\pi _i},{\bm {\pi}  _{-i}})]
\end{array}
\end{equation}
where the expectation of the immediate reward is defined as
\begin{equation}\label{im-reward}
\begin{array}{l}
E[{r_i}({s_i},{\pi _i(s_i)},{\bm {\pi} _{-i}(s_i)})]\\
=\sum\limits_{(a_i, \mathbf{a}_{-i}) \in \mathcal {A}_i} r_i(s_i,a_i, \mathbf{a}_{-i})\prod\limits_{z \in U} \pi_z(s_z,a_z)
\end{array}
\end{equation}
where $\pi_z(s_z,a_z)$ denotes the probability of the end user $u_z$ choosing action $a_z$ in state $s_z$. Moreover, the expectation of the discounted value of successor state is calculated with state transition probabilities
\begin{equation}\label{value-recursive}
\begin{array}{l}
E[ {V_i}({s_i^\prime},{\pi _i},{\bm {\pi}  _{-i}})]
=\sum\limits_{s_i^\prime \in \mathcal {S}_i} \mathbb{P}_{s_is_i^\prime}({\pi _i(s_i)},{\bm {\pi}  _{-i}(s_i)}){V_i}({s_i^\prime},{\pi _i},{\bm {\pi}  _{-i}})
\end{array}
\end{equation}
Therefore, in state $s_i(k)=s_i$, the end user takes action $a_i(k)$ and then gets the expectation value of the cumulative return under policy $\pi$ defined as state-action value function:
\begin{equation}\label{Q-value}
\begin{array}{l}
Q_i^{\pi}(s_i,a_i)\\
= E[\sum\limits_{\tau  = 0}^{ + \infty } {{\lambda ^\tau }{r_i}(s_i,a_i,\bm{\pi}_{-i}(s_i) )} \mid s_i(k)=s_i,a_i(k)=a_i]\\
=E[{r_i}(s_i,a_i,\bm{\pi}_{-i}(s_i) )+\lambda Q_i^{\pi}(s_i^\prime,a_i^\prime) \mid s_i(k)=s_i,a_i(k)=a_i]
\end{array}
\end{equation}
similarly, the first part of the Q-value function presents the current return of the end user by taking action $a_i$ in state $s_i$, which is defined as
\begin{equation}\label{Q-current-reward}
\begin{array}{l}
E[{r_i}({s_i},{a _i},{\bm {\pi} _{-i}(s_i)})]\\
=\sum\limits_{ \mathbf{a}_{-i} \in \mathcal {A}_{-i}} r_i(s_i,a_i, \mathbf{a}_{-i})\prod\limits_{z \in U \backslash \{i\} } \pi_z(s_z,a_z)
\end{array}
\end{equation}
Hence, (\ref{Q-value}) is reformulated as
\begin{equation}\label{re-Q-value}
\begin{array}{l}
Q_i^{\pi}(s_i,a_i)
=E[{r_i}({s_i},{a _i},{\bm {\pi} _{-i}(s_i)})] \\
\;\;\;\;\;\;\;\;\;\;\;\;\;\;\;+ \lambda\sum\limits_{s_i^\prime \in \mathcal {S}_i} \mathbb{P}_{s_is_i^\prime}({a _i},{\bm {\pi}  _{-i}(s_i)}){V_i}({s_i^\prime},{\pi _i},{\bm {\pi}  _{-i}})\\
\;\;\;\;\;\;\;\;\;\;\;\;\;\;\;=E[{r_i}({s_i},{a _i},{\bm {\pi} _{-i}(s_i)})] \\
\;\;\;\;\;\;\;\;\;\;\;\;\;\;\;+\lambda \sum\limits_{s_i^\prime \in \mathcal {S}_i} \mathbb{P}_{s_is_i^\prime}({a _i},{\bm {\pi}  _{-i}(s_i)})
\sum\limits_{a_i^\prime} {\pi_i (s_i^\prime,a_i^\prime)} Q_i^{\pi}(s_i^\prime,a_i^\prime)
\end{array}
\end{equation}

As discussed in Section III, there always exists an NE in our formulated
computation offloading game. Hence, the optimal strategy satisfies the Bellman's optimality equation as

\begin{equation}\label{Bellman_optimal}
\begin{array}{l}
{V_i}({s_i},{\pi^*_i},{\bm {\pi}^*_{-i}})
=\mathop {\max }\limits_{a_i \in \mathcal {A}_i} \{E[{r_i}({s_i},a_i,{\bm {\pi^*} _{-i}(s_i)})]\\
+\lambda \sum\limits_{s_i^\prime \in \mathcal {S}_i} \mathbb{P}_{s_is_i^\prime}({a _i},{\bm {\pi^*}  _{-i}(s_i)}){V_i}({s_i^\prime},{\pi^* _i},{\bm {\pi^*}_{-i}})\}
\end{array}
\end{equation}

Similarly, the optimal Q-value function $Q^*$ is the maximal action-value function over all the possible strategies, that is when all the end users follow the NE strategies given by

\begin{equation}\label{Q_optimal}
\begin{array}{l}
{Q_i}^*({s_i},a_i)
=E[{r_i}({s_i},a_i,{\bm {\pi^*} _{-i}(s_i)})]\\
+\lambda \sum\limits_{s_i^\prime \in \mathcal {S}_i} \mathbb{P}_{s_is_i^\prime}({a _i},{\bm {\pi^*}  _{-i}(s_i)}){V_i}({s_i^\prime},{\pi^* _i},{\bm {\pi^*}_{-i}})
\end{array}
\end{equation}

By combining (\ref{Bellman_optimal}) and (\ref{Q_optimal}), the optimal Q function is reformulated as

\begin{equation}\label{Q_optimal-new}
\begin{array}{l}
{Q_i}^*({s_i},a_i)
=E[{r_i}({s_i},a_i,{\bm {\pi^*} _{-i}(s_i)})]\\
+\lambda \sum\limits_{s_i^\prime \in \mathcal {S}_i} \mathbb{P}_{s_is_i^\prime}({a _i},{\bm {\pi^*}  _{-i}(s_i)}) \mathop {\max }\limits_{a_i^{\prime} \in \mathcal {A}_i} Q_i^*(s_i^{\prime}, a_i^{\prime})
\end{array}
\end{equation}

From (\ref{Q_optimal-new}), the optimal strategy means each end user $u_i$ chooses the optimal action which maximizes the corresponding Q-value function for current state, which forms an optimal policy over each time step. Moreover, the optimal Q function is determined by the joint action of all the end users and the joint policy, which makes it difficult to obtain the optimal strategy. In this paper, independent learning is used to solve the formulated computation offloading game. We consider each end user as an independent learner to learn its individual strategy without considering the actions taken by other end users, that is, for the end user, the other users are just one part of the environment. Actually, it's more practical because it's hard for the end user to be aware of the existence of the other end users, or to reduce complexity, it may choose to ignore the action information from other end users.

\subsection{Multi-Agent Q-learning for Computation Offloading}
In this section, Q-learning is applied to solve the independent MDP problems and an IL based MA-Q learning algorithm is proposed to solve the computation offloading with resource allocation problem in the multiuser IoT edge computing network. In the proposed algorithm, each end user runs an independent Q-learning algorithm and simultaneously learns an individual optimal strategy for their MDPs. Specifically, the selection of an optimal action depends on the Q-function, by following (\ref{Q_optimal-new}), the optimal Q-function is defined as the optimal expected value of state $s_i$ when taking action $a_i$ and then proceeding optimally. Since the state transition probability $\mathbb{P}_{s_is_i^\prime}$ and reward function $r_{s_is_i^\prime}$ are hard and impossible to be obtained in practice, mean return with multiple sampling is used to approximately indicate the expected cumulative reward. This is achieved by using Monte-Carlo (MC) Learning method, with sampling the same Q-function $Q_i(s_i,a_i)$ over different strategies. However, MC learning is complicated by calculating mean return with sampling complete episodes, so temporal difference (TD) learning is used to recursively update Q-value function with learning their estimates on the basis of other estimates, which is presented as
\begin{equation}\label{Q_function_update}
Q_i^\prime ({s_i},{a_i}) \leftarrow Q_i({s_i},{a_i}) + \alpha (r_i^{\prime} + \lambda  \mathop {\min }\limits_{a_i^{\prime}\in \mathcal{A}_i} Q({s_i^{\prime}},a_i^{\prime})-Q_i({s_i},{a_i})),
\end{equation}
where $r_i^{\prime} + \lambda \mathop {\min }\limits_{a_i^{\prime}\in \mathcal {A}_i} Q({s_i^{\prime}},a_i^{\prime})$ indicates the optimal cumulative returns at time slot $k+1$, which is called TD target. $\alpha$ is the learning rate $(0<\alpha \leq 1)$, for ensuring the convergence of Q-learning, the learning rate $\alpha_k$ is set as
\begin{equation}\label{alpha_k}
\alpha_k=\alpha_{k-1}*(\frac{{{\alpha _{end}}}}{{{\alpha _{ini}}}})^\frac{1}{{episodes}}
\end{equation}
where $\alpha_{ini}, \alpha_{end}$ are the given initial and last values of $\alpha$, respectively, and $episodes$ is the maximum iterations of the learning algorithm.

In this paper, an IL based MA-Q learning algorithm is proposed to address resource allocation problem in computation offloading for IoT edge computing networks. In the proposed algorithm, each end user runs a Q-learning procedure independently, and it only maintains its own Q table. The detailed process of the proposed algorithm is shown in $\textbf{Algorithm \ref{MAQ-learning}}$.

In the proposed algorithm, we use $\epsilon$-greedy method as the strategy of action selection, which focuses on solving the important problem of reinforcement learning, the exploration and exploitation trade-off. This gives a guide that the agent reinforces the best decision given information or explore new actions to gather more information. With the $\epsilon$-greedy method, the agent selects the optimal action corresponding to the largest Q-function with probability $1-\epsilon$, and chooses a random action with probability $\epsilon \in [0,1]$.

\begin{algorithm}
\renewcommand{\algorithmicrequire}{\textbf{Initialization:}}
\renewcommand{\algorithmicensure}{\textbf{Learning:}}
\caption{IL based MA-Q learning algorithm for computation offloading}
\label{MAQ-learning}
\begin{algorithmic}[1]
\REQUIRE~\\
Initialize parameters: discount factor $\lambda$, learning rate parameters $\alpha_{ini}, \alpha_{end}$, exploration rate $\epsilon$.\\
Set $k:=0$.
\FOR {$i$ to $U$}
\STATE Initialize action-value function $Q_i^k(s_i,a_i)$ \\
\STATE Initialize the resource allocation strategy $\pi_i^k(s_i,a_i)$\\
\STATE Initialize the state $s_i=s_i^k$
\ENDFOR
\ENSURE~\\
\WHILE {$k\leq K $}
\FOR {$i$ to $U$}
\STATE Choose an action $a_i$ according to the strategy $\pi_i(s_i)$
\STATE Measure the received SINR $\gamma_{i,j}^{m,n}$ at the receiver and the computation capacity of the gateway by identifying the transmit power, radio access technology and sub-channel.
\STATE Observe the current state $s_i^{k+1}$
\STATE Obtain a reward $r_i^k$ according to the measured information
\STATE Update $Q_i^{k+1}({s_i},{a_i})$ by (\ref{Q_function_update})
\STATE Update the strategy $\pi_i^{k+1}(s_i,a_i)$ according to $\epsilon$-greedy method
\STATE Set $k=k+1$ and $s_i=s_i^{k+1}$
\ENDFOR
\ENDWHILE
\end{algorithmic}
\end{algorithm}

\subsection{Convergence Performance of the Proposed Algorithm }
In practice, the requirement for Q-learning to obtain the correct convergence is that all the state action pairs $Q(s,a)$ continue to be updated.
Moreover, if we explore the policy infinitely, Q-value $Q(s,a)$ has been validated to converge with possibility 1 to $Q^*(s,a)$ , which is given by
\begin{equation}\label{1_possibility}
\mathop {\lim }\limits_{n \to \infty }\mathbb P_r(\left| {Q^*(s,a) - Q(s,a)_y} \right| \ge \varsigma ) = 0,
\end{equation}
where $y$ is the index of the obtained sample, and $Q^*(s,a)$ is the optimal Q-value while $Q(s,a)_y$ is one of the obtained samples. Therefore, Q-learning can identify an optimal action selection policy based on infinite exploration time and a partly-random policy for a finite MDP model. In this paper, we approximate the state and action space into finite states, and we use Monte-Carlo simulation to explore the possible policy, so we can obtain a near-optimal policy.

\section{Simulation Results}
\begin{table}
\centering
\caption{Simulation Parameters}
\begin{tabular}{|c|c|}
\hline
\multicolumn{2}{|c|}{Learning parameters}\\
\hline
 $ \alpha_{ini}, \alpha_{end}$ &  0.9, 0.001 \\
\hline
$ \lambda, \beta$ &  0.9, 5\\
\hline
\multicolumn{2}{|c|}{Channel parameters}\\
\hline
 $B^L, \sigma^L$ & $ 1.25*10^5Hz, -174+10log10B_L$\\
\hline
 $ B^W \sigma^W$ & $ 5*10^6Hz, -160+10log10B_W$\\
\hline
$\bar \gamma^L, \bar \gamma^W$ & $-15dB, 10dB$\\
\hline
$CH_L, CH_W$ & $ 10, 13$\\
\hline
$P_L, P_W$ & $\{0.05-0.3W\}$, $\{0.1-1W\}$ \\
\hline
\multicolumn{2}{|c|}{Computation parameters}\\
\hline
$d_{i,j}$ &$ \{10Kbits--4Mbits\}$\\
\hline
$\nu$& $ 500 \;cycles/bit$\\
\hline
$e^E, e^L$ & $10^{-7}, 10^{-8}\; W \;per \;CPU \;cycles $\\
\hline
$f$, $f_i$ & $10GHz, \{500MHz-1GHz \}$\\
\hline

\end{tabular}
\end{table}

In this section, the performance of our proposed IL based MA-Q learning algorithm for computation offloading with resource allocation problem is verified by simulations. In the considered IoT network, each end user runs a Q-table and independently interacts with the environment to learn its own optimal computation offloading policy. The distances between end users and the gateway are following normal distribution with $\mu=1000, \sigma=3$, that is, the users are located around a circle with its radius as $r=1km$. The Rayleigh fading is set as the small scale fading between users and the gateway with the parameter $B=3$, and the path-loss parameter $a=2.5$. We consider two radio access technologies, WiFi and LoRa. The detailed simulation parameters are given in Table 1.
\begin{figure}[!t]
  \centering
  \includegraphics[width=3.8in]{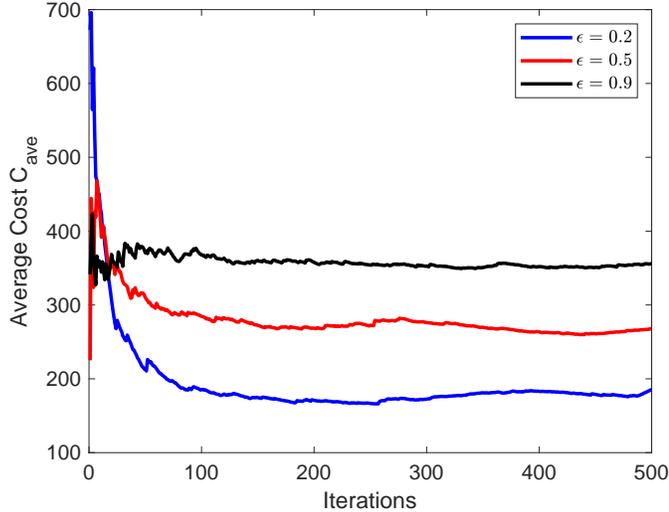}\\
  \caption{Convergence performance of the proposed IL-based MA-Q learning algorithm measured by average Cost $C_{ave}$, achieved by $u_1$, $Ag=30$. }\label{ave_epsilon}
\end{figure}

\begin{figure}[!t]
  \centering
  \includegraphics[width=3.8in]{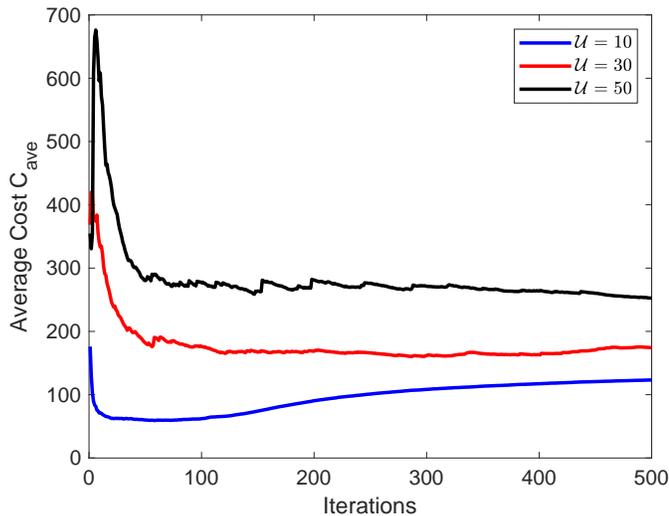}\\
  \caption{Performance comparison with different agent numbers $\mathcal{U}$ measured by average Cost $C_{ave}$, achieved by $u_1$, $\epsilon=0.2$.  }\label{ave_agents}
\end{figure}

Here, the maximum iteration episode for all the simulated curves is set as $episodes=10000$.
To verify the proposed IL-based MA-Q learning algorithm, without loss of generality, we consider an end user $u_1$ as an example. As shown in Fig.\ref{ave_epsilon}, it illustrates the convergence performance of the proposed algorithm under different exploration rates $\epsilon$. The average cost per time slot is converged with the iterations increasing, which indicates the Q-table of user $u_1$ has been trained stably. Moreover, we can see that the larger $\epsilon=0.9$ causes worse average cost, that is, the user explores too many random actions instead of exploiting the optimal action. From Fig.\ref{ave_epsilon}, the user has to pay more attention to exploiting its optimal action in the considered scenario.

\begin{figure}[!t]
  \centering
  \includegraphics[width=4in]{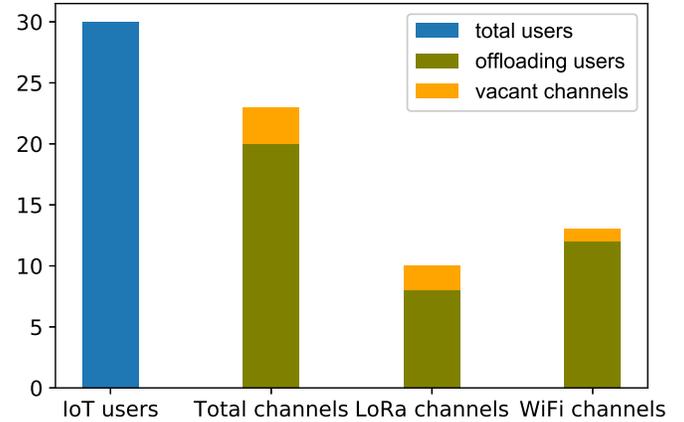}\\
  \caption{Access channels allocation with LoRa and WiFi access technologies, $Ag=30$. }\label{channels}
\end{figure}

\begin{figure}[!t]
  \centering
  \includegraphics[width=3.8in]{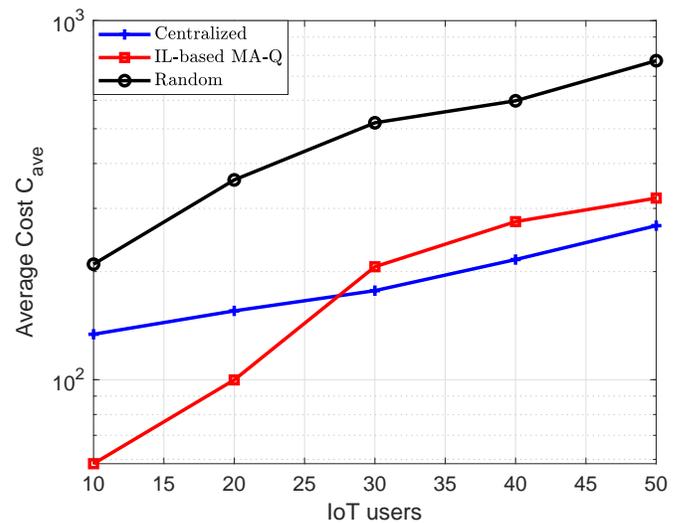}\\
  \caption{Comparisons for average cost $C_{ave}$, with different end users under algorithms Centralized, IL-based MA-Q, Random.}\label{comparison}
\end{figure}

Fig. \ref{ave_agents} shows convergence performance comparison with different number of agents. It's observed that the average cost, $C_{ave}$, per time slot of user $u_1$ is higher with more users in the considered IoT network. This is because with more users, it's harder for each user to access to the gateway due to the limited number of channels. Furthermore, with the trained Q-tables, we test all the users to make its own computation offloading decisions simultaneously. We consider there are 30 end users in the IoT network with 23 channels including 10 LoRa channels and 13 WiFi channels. From Fig. \ref{channels}, 20 users choose to offload their computation tasks while the other 10 users choose to execute their computation tasks locally. Here, 8 users offload their computation tasks using the LoRa channels while the others access to the gateway with WiFi channels. We can also observe that the end users can access to the channels reasonably without any collisions.

In Fig. \ref{comparison}, we investigate the average cost per time slot per end user under different computation offloading algorithms: Centralized, IL-based MA-Q and Random. Here, Centralized and Random algorithm are proposed as two benchmark algorithms for our proposed algorithm.
\begin{enumerate}
\item Centralized computation offloading: first, the gateway makes channel estimation for each user with the reference signals to obtain their channel information and computation task sizes, then it allocates users for local computing, or offloading computation tasks with LoRa channels or WiFi channels.
\item IL-based MA-Q computation offloading: each user independently runs its own Q-table to learn the optimal computation offloading policy by interacting with the environment.
\item Random computation offloading: each user randomly chooses to offload its computation tasks using LoRa channels or WiFi channels, or locally execute its computation tasks.
\end{enumerate}
Fig. \ref{comparison} shows that the average cost is increasing with the increase of end users in the considered IoT network. Moreover, our proposed IL-based MA-Q algorithm has lowest average cost when the end users are less than 27, while it has a bit higher cost than the Centralized algorithm with more end users. However, the proposed algorithm can be achieved in a distributed way, which reduces the computation burden on the gateway, and it saves the extra cost for channel estimation. At the same time, it has much better average cost performance than the Random algorithm.

\section{Conclusions}
Towards addressing the computation offloading problem with contending resources, like transmit power, radio access technology and sub-channel, in IoT edge computing networks, we investigated it as a resource allocation problem and proposed a distributed multi-agent based computation offloading mechanism. Moreover, an independent learners based multi-agent Q-learning (IL-based MA-Q) algorithm was developed to solve this resource allocation problem. The proposed computation offloading scheme enabled each end user to independently learn their computation policy so that computation burden on the centralized gateway was reduced. The feasibility of the proposed scheme applied to different scale of IoT networks has been verified by simulation results. Compared to the other two benchmark algorithms, it avoided radio access collisions and achieved lower system cost, and improved computation capacity for the gateway with distributed learning.
\bibliographystyle{IEEEtran}
\bibliography{IEEEabrv,mybib}

\end{document}